\documentclass[journal,comsoc]{IEEEtran}
\usepackage[T1]{fontenc}
\usepackage{caption}
\usepackage{cite}
\usepackage{cite}
\usepackage[pdftex]{graphicx}
\usepackage{subfigure}
\usepackage{amsmath}
\usepackage[cmintegrals]{newtxmath}
\usepackage[caption=false,font=normalsize,labelfont=sf,textfont=sf]{subfig}
\usepackage{url}
\usepackage{color}
\begin{document}

\title{A Novel Probability Weighting Method To Fit Gaussian Functions}

\author{Wei Chen \\{JiangXi University of Finances and Economics}
	\thanks{This work was supported by the National	Natural Science Foundation of China under Grants 61761020 and 62066017.}
	\thanks{}
	\thanks{}}

\markboth{}%
{Shell \MakeLowercase{\textit{et al.}}: Bare Demo of IEEEtran.cls for IEEE Communications Society Journals}

\maketitle

\begin{abstract}
Gaussian functions are commonly used in different fields, many real signals can be modeled into such form. Research aiming to obtain a precise fitting result for these functions is very meaningful. This manuscript intends to introduce a new algorithm used to estimate the full parameters of the Gaussian-shaped function. It is basically a weighting method, starting from Caruana's method, while the selection of weighting factors is from the statistics view and based on the estimation of the confidence level for the samples. Tests designed for comparison with current similar methods have been conducted. The simulation results indicate a good performance for this new method, mainly in precision and robustness.
\end{abstract}


\IEEEpeerreviewmaketitle

\section{Introduction}

\IEEEPARstart{G}{aussian} functions are widely used to describe physical and mathematical phenomena due to the common randomness in many fields \cite{Guo}. White noise, taking as an example, is a perennial research topic in the communication system, and its distribution can be simplified to a Gaussian model based on the statistic of large scale data \cite{whitenoise}. Another case is the reflective frequency spectrum from the fiber Bragg gratings (FBGs) when superluminescent light-emitting diodes (SLEDs) act as the source. Applying the proper apodization technique, such a reflective spectrum can be shaped perfectly into a Gaussian function \cite{Chen}. In brief, since the frequent usage of such type of function, it is mandatory and meaningful to study the Gaussian fitting approaches.

Currently, to achieve such a purpose, some algorithms have been proposed \cite{tot,Gf1,Gf2,Centroid}. The most common one is the so-called "Caruana's method" \cite{Caruana}. In this method, a logarithmic operation will be done to the deviation between acquired data and its corresponding true value. In such a way, this problem has been converted into a typical fitting problem about a polynomial of order 2, which can be easily solved by the least square method (LSM). This method brings much convenience to deal with the fitting problem for a Gaussian function using sampling data, and provides a direct way to obtain the relevant parameters compared with the previous iterative methods such as the Levenberg-Marquardt (LM) algorithm \cite{LM}.

To explore the potential and improve the performance of this algorithm, Guo proposed a weighting method based on it \cite{Guo}. By multiplying the weighting factors, this approach converts the manipulated error approximately back to the original noise. Such a method tends to equalize effort from the noise by increasing the weight of points with high amplitude. Compared with Caruana's method, relevant tests obtained from Guo's method imply better precision in full-spectrum fitting and also ability in analyzing incomplete spectrum. However, in Guo's method, the selection of weighting factors is an approximative way. In fact, for samples with low amplitude, such weighting factor would deviate badly from the supposed value, thus induce error to the fitting results.

This paper proposes an innovative algorithm to evaluate complete parameters for a Gaussian function. It is also based on Caruana's method, and improved from Guo's method in the determination of weighting factors. By a hypothesis test, the confidence level for each operated error is assigned as the corresponding weighting factor. In such a way, the weighting factors are more accurate and with higher solidarity in principal. As a consequence, this algorithm shows good performance in fitting Gaussian functions with the complete spectrum. Furthermore, after some modifications, it can be iterative, thus capable of achieving higher accuracy, and also features with excellent robustness during the numeric computation procedure.

\section{State Of Art of current research}
To obtain the full parameters of the Gaussian function, which are peak position $x_p$, width parameter $\sigma$, and peak amplitude $A$, the current commonly used method is Caruana's method \cite{Caruana}.  Its basic technique is to do a natural logarithm operation to the Gaussian function, leading this problem transferred to a simple polynomial fitting problem.\\
\begin{equation} 
\begin{split}
\ln(y)&=\ln(Ae^{-\frac{(x-x_p)^2}{2\sigma^2}})\\
&=\ln(A)+(-\frac{x^2}{2\sigma^2})+\frac{x_p}{\sigma^2}x+\frac{x_p^2}{2\sigma^2} 
\end{split}
\label{Gaussianfunction}
\end{equation} 
Set $a=\ln(A)+\frac{x_p^2}{2\sigma^2}$, $b=\frac{x_p}{\sigma^2}$, $c=-\frac{1}{2\sigma^2}$, then $\ln(y)=a+bx+cx^2$. Consider noise $\varepsilon$ between the received signal $\bar{y}$ and the true one $y$,
 \begin{equation} 
 \varepsilon =\bar{y}- y \label{error0}
 \end{equation} 
In Caruana's method, the error after logarithm operation $\delta$ is analyzed, that is
 \begin{equation} 
 \begin{split}
 \delta &=\ln \bar{y}-\ln y \\
        &=\ln \bar{y}-(a+bx+cx^2) \label{error1}
  \end{split}      
 \end{equation} 
Following the LSM, consider the expectation of $\delta^2$,
 \begin{equation} 
 \begin{split}
 \mathbb{E}\{\delta^2\} &=(\ln \bar{y}-\ln y )^2\\
 &=(\ln \bar{y}-(a+bx+cx^2))^2 \label{error2}
 \end{split}  
 \end{equation} 
To minimize the expectation, do the partial differential with respect to $a$, $b$ and $c$, this problem comes to solve such linear system:
\begin{equation}
\begin{bmatrix}
\ N & \sum_{i=1}^N x_i & \sum_{i=1}^N x_i^2\\
\sum_{i=1}^N x_i & \sum_{i=1}^N x_i^2 &\sum_{i=1}^N x_i^3 \\
\sum_{i=1}^N x_i^2 & \sum_{i=1}^N x_i^3 &\sum_{i=1}^N x_i^4 \\
\end{bmatrix}\begin{bmatrix}
a\\b\\c
\end{bmatrix}=\begin{bmatrix}
\sum_{i=1}^N \ln\bar{y}_i\\
\sum_{i=1}^N x_i\ln\bar{y}_i\\
\sum_{i=1}^N x_i^2\ln\bar{y}_i
\end{bmatrix} \label{linearsystemcaruana}
\end{equation}
N is the number of samples selected. The main idea of Caruana's method is to change a complex nonlinear problem to a simple linear one by a logarithmic operation, which greatly increases the effectiveness. However, this method also emerges some disadvantages, such as high sensitivity to the noise and large fluctuation of the results from repeating tests.

To lower such effects, Guo proposed a new weighting method based on Caruana's method, declaring that it can increase the fitting accuracy and noise sensitivity \cite{Guo}. Basically, it is a weighting method, setting the sample amplitudes as corresponding weighting factors. By multiplying such weighting factors, it tried to transfer the error after logarithmic operation back to white Gaussian noise as \eqref{GuoError1} and  \eqref{GuoError2} shown.
\begin{equation} 
\begin{split}
y\delta &=y(\ln \bar{y}-\ln y)\\&=y(\ln(y+\varepsilon)-\ln y)
\end{split}
\label{GuoError1}
\end{equation} 
When the amplitudes of selected points are far much higher than the noise level, the weighted error can be approximated to
\begin{equation} 
\begin{split}
y\delta &=y(\ln(y+\varepsilon)-(a+bx+cx^2))\\
&\approx y(\ln y-(a+bx+cx^2))+\varepsilon 
\end{split}
\label{GuoError2}
\end{equation} 
We can see that the latter term $\varepsilon$ is the background white noise, independent with $a, b$ and $c$. So the purpose is to optimize the square sum of the first term, which is a simple fitting problem of a polynomial with order 2. Since $y$ is unknown, replace it with $\bar{y}$, the corresponding linear system is
\begin{equation}
\begin{bmatrix}
\sum_{i=1}^N \bar{y_i}^2 & \sum_{i=1}^N x_i{\bar{y_i}^2} & \sum_{i=1}^N x_i^2{\bar{y_i}^2}\\
\sum_{i=1}^N x_i{\bar{y_i}^2} & \sum_{i=1}^N x_i^2{\bar{y_i}^2} &\sum_{i=1}^N x_i^3{\bar{y_i}^2} \\
\sum_{i=1}^N x_i^2{\bar{y_i}^2} & \sum_{i=1}^N x_i^3{\bar{y_i}^2} &\sum_{i=1}^N x_i^4{\bar{y_i}^2} \\
\end{bmatrix}\begin{bmatrix}
a\\b\\c
\end{bmatrix}=\begin{bmatrix}
\sum_{i=1}^N  \bar{y_i}^2\ln\bar{y}_i\\
\sum_{i=1}^N x_i  \bar{y_i}^2\ln\bar{y}_i\\
\sum_{i=1}^N x_i^2 \bar{y_i}^2\ln\bar{y}_i
\end{bmatrix} \label{linearsystemGuo}
\end{equation}

Obviously, Guo's method tends to uniformize the errors between received samples and respective true values by multiplying the weighting factors. From the simulation, it shows rather big advantages in fitting the Gaussian function compared with Caruana's method. It features with higher precision and accuracy, also good resistivity to the noise.

However, since this method did an approximation in \eqref{GuoError2} under the condition of high sample amplitude, it is sensitive in point selection by nature. If the spectrum analyzed contains samples not around the peak, such approximation operation would induce large error.
\section{Proposed new algorithms}
In this manuscript, we submit a new algorithm trying to improve the quality of Gaussian function fitting. It is also a weighting method, similar to Guo's method. However, the setting of weighting factors is based on probability theory, thus we find that Guo's method is a special case of ours. It is simple, fast, and with fine performance in parameter estimations for Gaussian functions.

Also consider the error after logarithmic operation in \eqref{error1} and do some tricks, it can be rewritten to
 \begin{equation} 
\delta= \ln(1+\frac{\varepsilon}{y}) \label{error3}
 \end{equation} 
Obviously, the distribution of $\delta$ is no longer the same as original noise  $\varepsilon$ but changes with $y$. Noting the probability density function (PDF) of the noise $\varepsilon$ as $f_n(x)$, it is easy to obtain the PDF of $\delta$
\begin{equation}
f_\delta(x)=ye^xf_n(y(e^x-1)) \label{fd}
\end{equation}

From the view of the hypothesis test, to estimate the acceptance probability for each sample, we can set a uniform threshold for the error $\delta$, noted as $M$. Received samples with bias beyond the confidence interval $[-M, M]$ would be refused. Based on the provided PDF in \eqref{fd}, we can calculate the corresponding acceptance probability or confidence level for each sample, written as
\begin{equation}
P_i=\int_{-M}^{M} f_{\delta_i}(x) dx \label{pi1}
\end{equation}
The value $P_i$ indicates the confidence level of received $y_i$ under the assumed confidence interval $[-M, M]$. Thus it is reasonable to choose $P_i$ as the weighting factor in an improved LSM to determine the parameters of the Gaussian function. By multiplying such weighting factors, we equalize the contribution of each sample deviation $\delta_i$ to the sum. This criterion of setting weighting factors is called "Uniform Confidence Principle" by us, which can be generalized and extended to similar optimization issues. In fact, for a common and original fitting problem in the least square sense, each error is purely raw background noise and shares the same distribution function, thus corresponding confidence levels are all the same. In our case, error after weighting is shown as
    \begin{equation}    
   \varepsilon_i=P_i\delta_i =P_i(\ln \bar{y}-(a+bx+cx^2)) \label{pi3}      
    \end{equation}
Conducting the partial differential to the sum of $\varepsilon_i^2$ according to LSM, the problem is transferred to solve such linear system below:
\begin{equation}
\begin{bmatrix}
\sum_{i=1}^N P_i^2 & \sum_{i=1}^N x_i{P_i^2} & \sum_{i=1}^N x_i^2{P_i^2}\\
\sum_{i=1}^N x_i{P_i^2} & \sum_{i=1}^N x_i^2{P_i^2} &\sum_{i=1}^N x_i^3{P_i^2} \\
\sum_{i=1}^N x_i^2{P_i^2} & \sum_{i=1}^N x_i^3{P_i^2} &\sum_{i=1}^N x_i^4{P_i^2} \\
\end{bmatrix}\begin{bmatrix}
a\\b\\c
\end{bmatrix}=\begin{bmatrix}
\sum_{i=1}^N P_i^2\ln\bar{y}_i\\
\sum_{i=1}^N x_i P_i^2\ln\bar{y}_i\\
\sum_{i=1}^N x_i^2 P_i^2\ln\bar{y}_i
\end{bmatrix} \label{linearsystem}
\end{equation}
It is common to suppose the additional noise is white, with mean zero and variance $\sigma_n^2$. Without losing generality, we also make this assumption. Therefore, $P_i$ can be calculated
\begin{equation}
P_i= \Phi(\frac{y_i(e^{M}-1)}{\sigma_n})- \Phi(\frac{y_i(e^{-M}-1)}{\sigma_n})
\end{equation}
$\Phi$ is the cumulative probability function of standard normal distribution. In Gaussian function fitting, it is reasonable to choose points with high amplitude to lower the effect from the noise, which implies $\delta$ is close to zero in practice. So, the threshold value $M$ also is small under a proper assumption of confidence level(e.g.,95\%). As a consequence, $P_i$ is approximately expressed as
\begin{equation}
P_i \approx 2\Phi(\frac{y_iM}{\sigma_n})-1
\end{equation}

Furthermore, making a linear approximation to simplify calculation or under the assumption of uniform distribution for background noise, confidence level $P$ will be linear with $y$ instead. As a result, for each sample, the corresponding weighting factor $P_i$ can be approximated to $y_i/y_p$, with $y_p$ is the peak amplitude. Since scaling operation about the weighting factors in  \eqref{linearsystem} does not affect the result, $P_i$ can be rewritten as
\begin{equation}
P_i\approx y_i
\end{equation}
It is just the weighting factor used in Guo's method.

From the previous operations, it can be concluded that Guo's method is an approximation or a specific case of the proposed weighting algorithm. The novel method shows a higher level of generality in principle and should be more capable to handle various noisy environments. However, this method requires prior statistical information about the noise, which is always not difficult to get in practice, at least some simplified and empirical models are available.

\section{Simulation and comparison}
In this section, we do simulation for all three algorithms mentioned above, which are Caruana’s method, Guo's method, and the novel probability weighting method. Then we conduct the comparison for simulation results, mainly focusing on the accuracy and precision of obtained parameters. The robustness of these algorithms to noise will be studied by modifying the SNR. Furthermore, the ability to handle incomplete spectrum also has been investigated in this section, by iterating the two weighting methods.  

During our experiment, without loss of generality, we select peak position $x_p=5$, standard deviation $\sigma=0.2$, and peak amplitude $A=1$. Firstly, the sampling rate is set to 10 to check the feasibility of these algorithms. And we assume that the noise is white with zero mean and a standard deviation $\sigma_n$ as mentioned before.

For signals with a complete spectrum, about the selection of points, we set a threshold firstly, then starting from the peak and moving bidirectionally until the first point with amplitude lower than the threshold, samples in such interval are chosen. In our case, the threshold value is set to $2\sigma_n$ empirically. 

About the new method, for peak point, we suppose the background noise with a confidence level $95\%$, that indicates an interval about $(-2\sigma_n,2\sigma_n)$. Then corresponding confidence interval for error in the peak $\delta_p$ is about $(\frac{-2\sigma_n}{y_p},\frac{2\sigma_n}{y_p})$, noted $\frac{2\sigma_n}{y_p}$ as $M$. After the determination for $M$, respective weighting factor $P_i$ for each $y_i$ is derived consequently, written as $P_i \approx 2\Phi(\frac{2y_i}{y_p})-1$.

We conduct the simulation under SNR=14dB, 16.5dB, 20dB, 26dB, 32dB, 40dB, and 46dB, which means $\sigma_n=0.2$, 0.15, 0.1, 0.05, 0.025, 0.01, and 0.005 respectively, then repeat it for 5000 times under each noisy environment.

\begin{figure}
	\centering
	\subfigure[Peak position]{\includegraphics[width=2.5in]{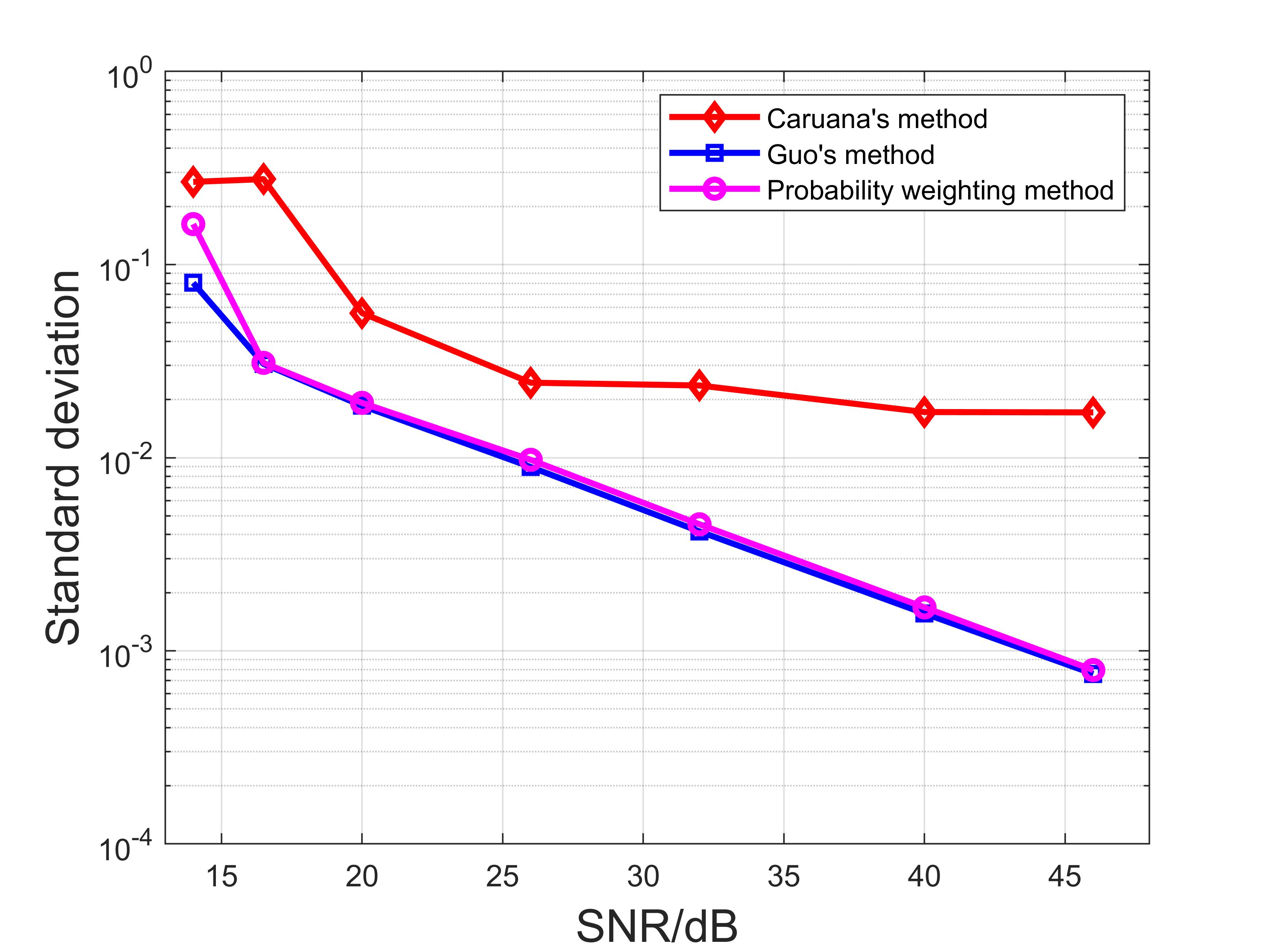}}
	\subfigure[$\sigma$]{\includegraphics[width=2.5in]{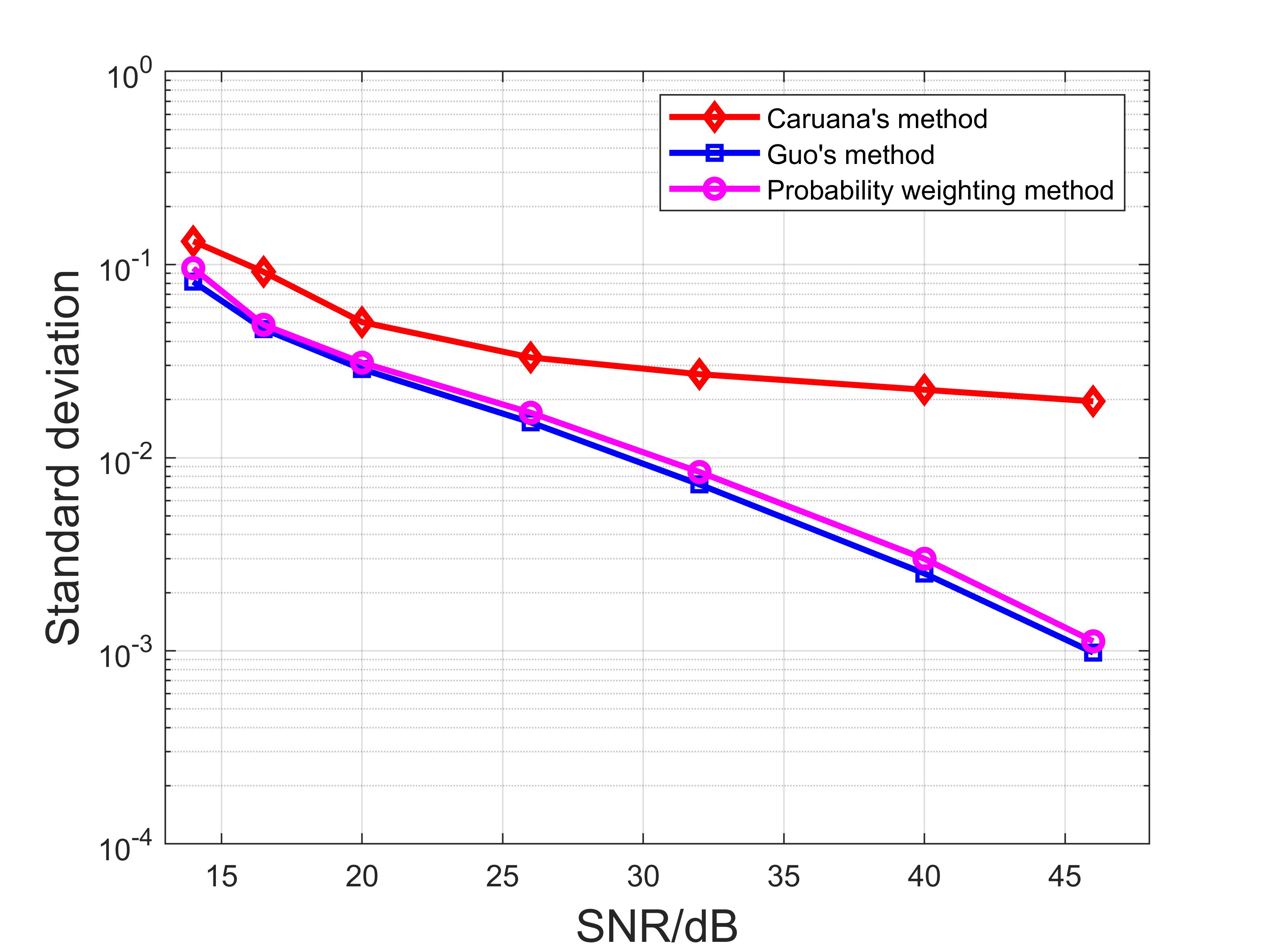}}
	\subfigure[Peak amplitdue]{\includegraphics[width=2.5in]{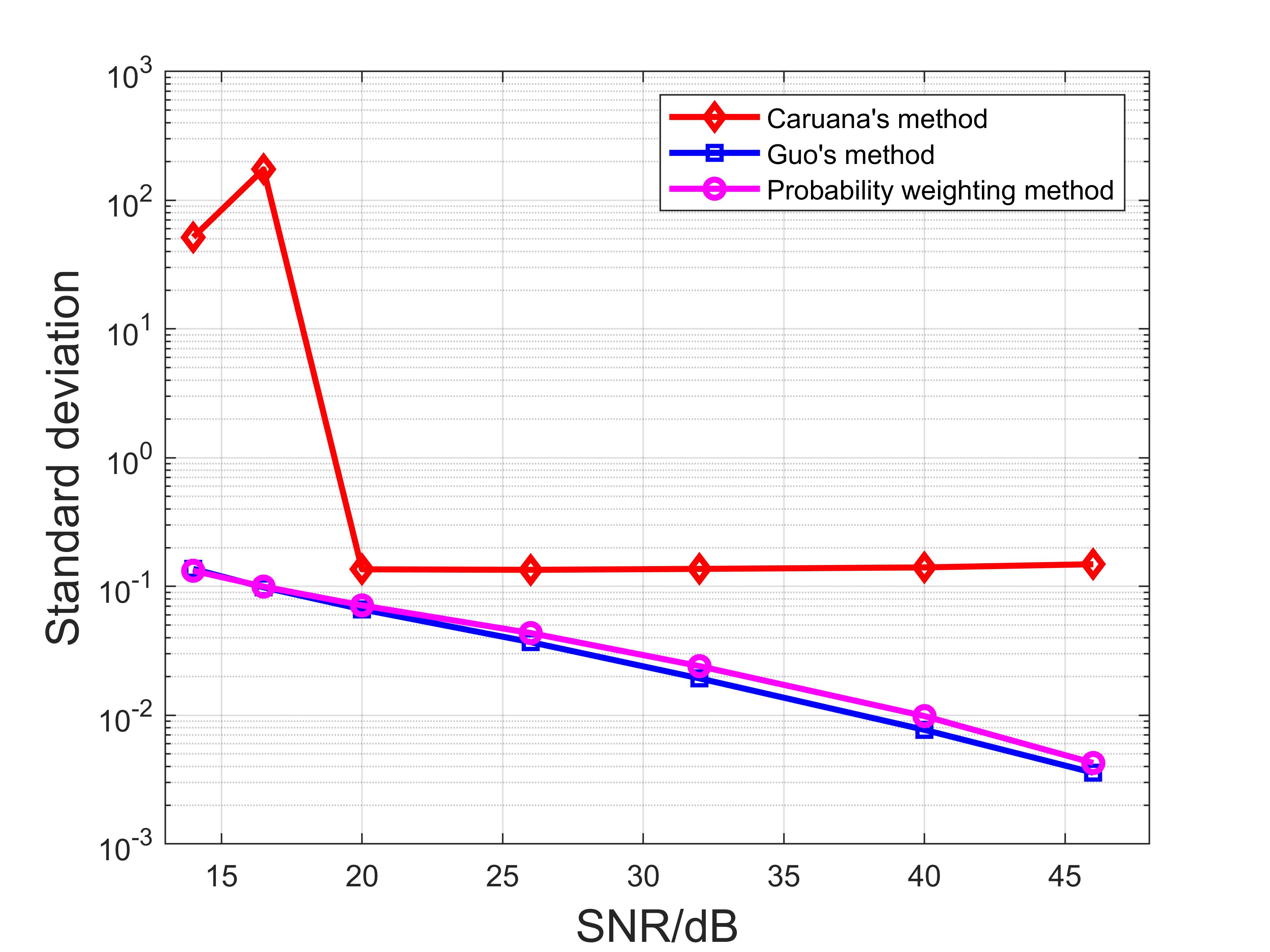}}
	\caption{Comparison of three methods at different SNR}
	\label{Comptot}
\end{figure}

From the results obtained, we find that the accuracy values of all three parameters are much smaller than those of corresponding precision, some even by several orders of magnitude. So in the determination of fitting quality, precision values dominate. Fig.\ref{Comptot} shows the standard deviations of the three parameters in different noise levels. We can see that Guo's method and the proposed new weighting method are almost with the same performance, and much better than that of Caruana's method, especially in high SNR environments. Certainly, precisions for all three parameters improve as the increasing of SNR, but less sensitive for Caruana's method. The precision of two weighting methods grossly shows a linear relationship with SNR in the semi-logarithmic scale.

However, the proposed new algorithm shows no significant advantage to Guo's method during the tests. One of the possible reasons is the sparseness of the spectrum used in the simulation. Few number of samples may induce a lack of statistical characters, thus cause the noise distribution deviates much from white assumption. The other reason is the approximation during the calculation, which causes the weighting factors differ from its true value.

\begin{figure}
	\centering
	\includegraphics[width=2.5in]{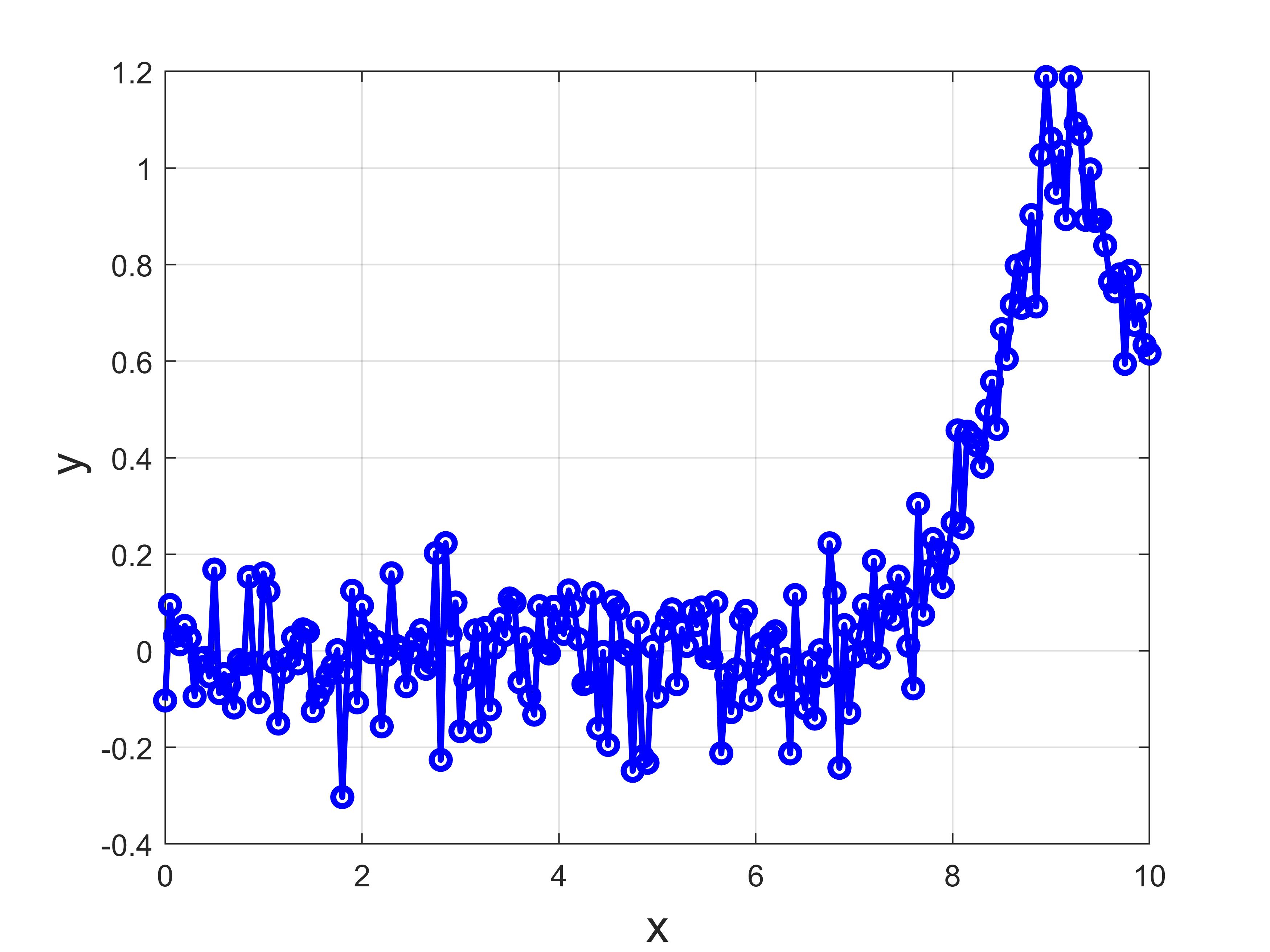}
	\caption{Gaussian signal with incomplete spectrum}
	\label{incomplete}
\end{figure}

Under some conditions, signal with complete spectrum is impossible, but always like Fig. \ref{incomplete} shown, being so-called long-tailed. To deal with this problem, in principle, we can use the former three methods by solving a linear system. Since lack of samples with higher amplitude on the other side of the peak, information from the long tail must be adopted. However, this operation would introduce samples with low amplitude, which are totally dominated by noise. As a consequence, the fitting results would deviate greatly from the true values. To increase the accuracy, we observed that by their weighting nature, Guo's method and proposed probability method can be iterative by updating the corresponding weighting factors, which are both related to the sample amplitudes. Updating can be done by using previous fitting results during the iteration process. As a result, the linear system should be modified as
\begin{equation}
\begin{aligned}
\begin{bmatrix}
\sum_{i=1}^N {P_i^{(k)}}^{2} & \sum_{i=1}^N x_i{P_i^{(k)}}^{2} & \sum_{i=1}^N x_i^2{P_i^{(k)}}^{2}\\
\sum_{i=1}^N x_i{P_i^{(k)}}^{2} & \sum_{i=1}^N x_i^2{P_i^{(k)}}^{2} &\sum_{i=1}^N x_i^3{P_i^{(k)}}^{2} \\
\sum_{i=1}^N x_i^2{P_i^{(k)}}^{2} & \sum_{i=1}^N x_i^3{P_i^{(k)}}^{2} &\sum_{i=1}^N x_i^4{P_i^{(k)}}^{2}  
\end{bmatrix}     
\begin{bmatrix}
a^{(k)}\\b^{(k)}\\c^{(k)}
\end{bmatrix}\\
=\begin{bmatrix}
\sum_{i=1}^N {P_i^{(k)}}^{2}\ln\bar{y}_i\\
\sum_{i=1}^N x_i{P_i^{(k)}}^{2}\ln\bar{y}_i\\
\sum_{i=1}^N x_i^2{P_i^{(k)}}^{2}\ln\bar{y}_i
\end{bmatrix}
\end{aligned}
\label{Iterations}
\end{equation}
The weighting factor ${P_i^{(k)}}$ also can be approximated as $2\Phi(\frac{{y_i^{(k)}}M}{\sigma_n})-1$, where 
\begin{equation}
y_i^{(k)}=
\begin{cases}
\bar{y}_i& \text{ $k=0$ } \\
e^{a^{(k)}+b^{(k)}x+c^{(k)}x^2}& \text{ $k>0$ }
\end{cases}
\label{updating}
\end{equation}

To test the capability in handling the incomplete spectrum, we set a signal with peak position $x_p=9.2$, $\sigma=0.75$ and peak amplitude $A=1$, the noise level $\sigma_n$ equal to 0.1. The $x$ range is from zero to ten, such that there are few samples on the right remained.

\begin{figure}
 	\centering
 	\includegraphics[width=2.5in]{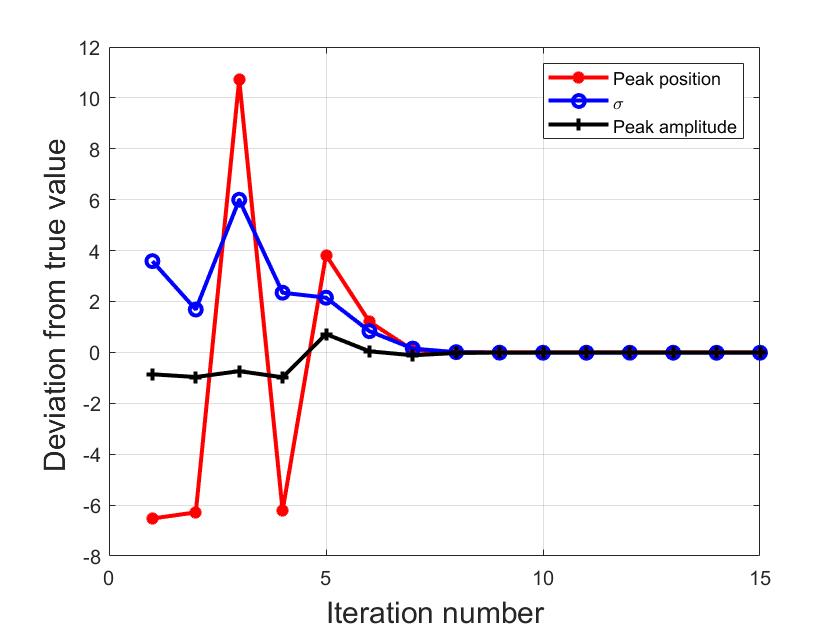}
 	\caption{Convergence of probability weighting method }
 	\label{IterOfMe}
 \end{figure}

In Fig. \ref{IterOfMe}, the convergence conditions for all three parameters obtained by the probability weighting method are illustrated. As we predicted, in the first iteration, there is a large error between the fitting results and corresponding true values. However, fortunately, as the iteration progresses, the deviations decrease quickly. After about 9 times of iterations, all results convergent to their limits, the absolute deviations are about 0.005, 0.01, and 0.007 for peak position, $\sigma$ and peak amplitude respectively. The convergences are fast and the accuracies are rather high but still significantly lower than those with a full spectrum and sample selection. During the simulations, the new probability weighting method shows fine robustness in numeric computing, no singularity appeared during the procedure of solving the linear system.

\section{Conclusion and Discussion}
In this paper, we introduce a novel probability weighting method to estimate the parameters of Gaussian functions. It is inspired by the hypothesis test. We consider each operated error instead of the original one and choose the acceptance probability as the corresponding weighting factors. In fact, this weighting factor determination criterion, named "Uniform Confidence Principle", can be extended to other similar fitting problems. If the target optimization function is the deviation not directly originating from the noise but after some mathematical operations, to increase the optimization quality, this novel method will be a promising option. It shows higher generality in principal, Guo's method and common LSM used in fitting problems can be viewed as special cases of this new method. And it also obtains comparably good results with Guo's method. Worthy to mention, the probability weighting method also can be iterative, such that a better performance in handling incomplete spectrum has been achieved. In this test, it shows fast convergence speed, improved accuracy, and fine numeric computation stability.


%


\ifCLASSOPTIONcaptionsoff
  \newpage
\fi












\bibliographystyle{ieeetr}

\bibliography{ref}

\end{document}